\documentclass[journal,twocolumn]{IEEEtran}
\usepackage{graphicx}
\usepackage{caption}
\usepackage{makecell}
\usepackage{soul}
\usepackage{listings}
\usepackage{multirow}
\usepackage{pifont}
\usepackage{xspace}
\usepackage{epigraph} 
\usepackage[flushleft]{threeparttable}
\usepackage{url}
\usepackage{algorithm2e}
\UseRawInputEncoding
\usepackage[colorinlistoftodos]{todonotes}
\usepackage{listings}
\usepackage{graphicx,wrapfig}
\usepackage{comment}
\usepackage{fontawesome}
\usepackage{booktabs}
\usepackage[strict]{changepage}
\usepackage{framed}
\usepackage{lipsum}
\usepackage{graphicx}
\usepackage{subfloat}
\usepackage[table]{xcolor}

 \usepackage[colorinlistoftodos]{todonotes}

\definecolor{chestnut}{rgb}{0.8, 0.36, 0.36}

\definecolor{chestnut}{rgb}{0.8, 0.36, 0.36}

\usepackage[skins,breakable]{tcolorbox}

\definecolor{Large}{HTML}{696969}
\definecolor{Negligible}{HTML}{D3D3D3}
\definecolor{Medium}{HTML}{808080}
\definecolor{Small}{HTML}{A9A9A9}
\definecolor{a1blue}{HTML}{FFE5CC}   
\definecolor{a1green}{HTML}{D0F0EF}  

\definecolor{c1purple}{HTML}{E0D5F7} 
\definecolor{c1pink}{HTML}{FFE0F0}   
\definecolor{B1blue}{HTML}{9ecae1}
\definecolor{A1blue}{HTML}{c6dbef}
\definecolor{C2blue}{HTML}{084594}
\definecolor{C1blue}{HTML}{2171b5}
\definecolor{lightgray2}{HTML}{f6f6f6}
\definecolor{chatgptgreen}{HTML}{bbe7dc}

\definecolor{ori}{HTML}{D3D3D3}  

\usepackage{tikz}
\usepackage{xparse}
\usepackage{xstring}
\usepackage{xcolor}
\usepackage[margin=1in]{geometry}
\usepackage{lmodern}
\usepackage{microtype}

\definecolor{userblue}{HTML}{D0E8FF}
\definecolor{botgreen}{HTML}{DFFFE0}
\definecolor{namegray}{gray}{0.4}
\definecolor{codegray}{gray}{0.95}

\newcommand{\RqOne}{\textbf{RQ1:} \emph{What is the baseline English proficiency (CEFR level) of problem descriptions generated by LLMs?}}
\newcommand{\RqTwo}{\textbf{RQ2:} \emph{Does the English proficiency (CEFR level) of a natural language prompt influence the code proficiency and correctness of the solutions generated by LLMs?}}

\begin{document}

\title{How Natural Language Proficiency Shapes GenAI Code for Software Engineering Tasks}

\author{
\IEEEauthorblockN{
Ruksit Rojpaisarnkit\IEEEauthorrefmark{1}, Youmei Fan\IEEEauthorrefmark{1}, 
Kenichi Matsumoto\IEEEauthorrefmark{1}}
 and
 Raula Gaikovina Kula\IEEEauthorrefmark{2}
\\
\IEEEauthorblockA{
\IEEEauthorrefmark{1}Nara Institute of Science and Technology, Japan,
\IEEEauthorrefmark{2}The University of Osaka, Japan,
\\
rojpaisarnkit.ruksit.rn1@is.naist.jp, fan.youmei.fs2@is.naist.jp,  matumoto@is.naist.jp, raula-k@ist.osaka-u.ac.jp}
}

\maketitle

\begin{abstract}
With the widespread adoption of Foundation Model (FM)-powered tools in software engineering, the natural language prompt has become a critical interface between developers and Large Language Models (LLMs). 
While much research has focused on prompt structure, the natural language proficiency is an underexplored factor that can influence the quality of generated code. 
This paper investigates whether the English language proficiency itself independent of the prompting technique
affects the proficiency and correctness of code generated by LLMs. 
Using the HumanEval dataset, we systematically varied the English proficiency of prompts from basic to advanced for 164 programming tasks and measured the resulting code proficiency and correctness. Our findings show that LLMs default to an intermediate (B2) natural language level. While the effect on the resulting code proficiency was model-dependent, we found that higher-proficiency prompts consistently yielded more correct code across all models. 
These results demonstrate that natural language proficiency is a key lever for controlling code generation, helping developers tailor AI output and improve the reliability of solutions.



\end{abstract}


\section{Introduction}
Foundation Model (FM)-powered coding assistants are disrupting the software development landscape. Developers are utilizing AI technologies like GitHub Copilot\footnote{https://github.com/features/copilot}, ChatGPT\footnote{https://openai.com/index/chatgpt/}, and Gemini\footnote{https://gemini.google.com/app} to generate code and solve software engineering tasks. In recent years, Large Language Models (LLMs) have become powerful coding assistants, particularly for code generation~\cite{jiang2024survey,jin2024can}. The efficacy of these models heavily relies on the natural language prompts they receive, making prompt quality a key determinant of output quality~\cite{liang2024large}.

However, while much research has focused on prompt engineering and model fine-tuning to improve results~\cite{pornprasit2024fine,shin2023prompt}, less attention has been paid to how the natural language complexity of the prompt itself affects the generated code. The proficiency of the generated code is a critical, yet overlooked factor. A mismatch between the code complexity and the user's skill level can create significant issues. This can lead to novices being overwhelmed by overly complex code or experts being frustrated by simplistic, inefficient solutions. Consider a developer who is highly proficient at describing tasks in natural language (e.g., C1-level proficiency) but has only a novice-level understanding of the programming language (e.g., A1-level proficiency). If a LLMs respond to the sophisticated prompt and generates code that is also at a C1-level proficiency, this developer may face significant challenges in understanding, debugging, and safely integrating the code. Recent studies show that students and practitioners often struggle to understand LLM-generated code due to unfamiliar code styles, an overestimation of its correctness, and their own limited programming proficiency~\cite{Zi_2025}. Furthermore, other research has found that beginning programmers struggle to write effective prompts and often misunderstand the resulting code, revealing a fundamental disconnect between novice users and LLMs~\cite{Nguyen_2024}.

\begin{table*}[]
\centering
\begin{tabular}{p{3cm}p{12cm}}
\toprule
\textbf{Prompt} & \textbf{Sample AI-Generated Response} \\
\midrule
Translate the following software engineering problem into natural language including function signature, docstring, body, and unit testcases without code. Make sure it fits into one paragraph.  & The function `has\_close\_elements' takes a list of floating-point numbers, `numbers', and a floating-point `threshold'. It checks if there are any two numbers in the list that are closer to each other than the specified threshold. The function returns `True' if such a pair of numbers exists, and `False' otherwise. For example, calling `has\_close\_elements' with the list `[1.0, 2.0, 3.0]' and a threshold of `0.5' returns `False' because no two numbers in the list are within `0.5' units of each other. Conversely, calling it with `[1.0, 2.8, 3.0, 4.0, 5.0, 2.0]' and a threshold of `0.3' returns `True' because at least one pair of numbers is closer than `0.3' units apart. \\

\bottomrule
\end{tabular}
\caption{AI-Generated Descriptions of Software Engineering Problems}
\label{tab:promptdesign}
\end{table*}

This challenge extends to various settings. In a collaborative industry environment, it can create significant friction and increase review overhead. For instance, if a senior developer uses LLMs to produce highly complex, idiomatic code, junior team members may struggle to understand, review, and later maintain it, impacting team velocity. Furthermore, modern software teams often operate in environments where developers have varied expertise across multiple languages. Ensuring AI-generated code aligns with their specific proficiency in the target language is crucial for preventing subtle bugs~\cite{10.1145/3631967}. Therefore, tailoring the proficiency of generated code to user expertise is a critical factor for the successful adoption of LLMs.

In this paper, we conduct an empirical study using several leading LLMs (GPT-4o, Gemini 2.5 Pro, and Claude Sonnet 4) and the HumanEval dataset~\cite{chen2021codex} to investigate whether the natural language proficiency of a prompt affects the proficiency and correctness of the generated code. Briefly, the proficiency of an artifact (a prompt or generated code) refers to the proficiency level required by a person to author or comprehend it. We modified the English proficiency level of each problem description based on the Common European Framework of Reference for Languages (CEFR), a robust international standard for defining language proficiency. 

By prompting we align the CEFR \cite{cefr2020} according to the following levels and example words:
\begin{itemize}
    \item CEFR A1 - Beginner (i.e. about, above)
    \item CEFR A2 - Elementary (i.e. ability, access, acceptance)
    \item CEFR B1 - Intermediate (i.e. accidental, accompany)
    \item CEFR B2 - Upper Intermediate (i.e. absurd, actual)
    \item CEFR C1 - Advanced (i.e.complexity, reluctantly)
    \item CEFR C2 - Proficient (i.e. eccentricity, deliberate)
\end{itemize}

Our study is based on the software engineering tasks dataset from HumanEval, which contains 164 hand-written programming problems in Python. Each problem includes a prompt (doc-string), a canonical solution, and test cases. To address known issues with the original test cases and ensure the reliability of our pass@1 metric, we incorporate the more robust test suite from HumanEvalPlus~\cite{chen2021codex}.

To measure code proficiency, we adapt the levels defined by PYCEFR~\cite{pycefr2022}, which assigns proficiency ratings to Python elements based on their conceptual difficulty. For example, \textit{print()} and \textit{return} are considered A1 (Beginner), \textit{if/else} and \textit{assert} are B1 (Intermediate), and \textit{zip()} and \textit{map()} are C2 (Proficient).

We then ask the following research questions:
\begin{enumerate}
    \item \RqOne \\
    While the challenges of code comprehension are known, the causal link from natural language prompt proficiency to code proficiency is not well understood. This research question aims to investigate the proficiency level of LLMs-generated descriptions based on the CEFR based proficiency. 
    
    \item \RqTwo \\
    With the baseline established from RQ1, we can then systematically explore the core of the problem. This research question investigates whether the natural language proficiency of problem descriptions influences the code proficiency of solutions generated by LLMs. 
\end{enumerate}

Our findings reveal that LLMs tend to generate problem descriptions at a B2 (Upper Intermediate) level or higher, suggesting a baseline English proficiency is required to understand their output. Furthermore, a clear correlation exists for some LLMs: adjusting the natural language proficiency of prompts tend to yield different proficiency levels of code. Accordingly, simplifying the natural language proficiency of the prompt can lower the correctness of the resulting code. These results indicate that the natural language proficiency of a prompt directly influences the reasoning and code generation capabilities of LLMs. The dataset of prompts and responses analyses for each RQ are available in Zenodo, at \url{https://doi.org/10.5281/zenodo.17285168}.

\begin{table*}[]
\centering
\begin{tabular}{p{3cm}p{12cm}}
\toprule
 \textbf{Prompt} & \textbf{Sample AI-Generated Response} \\
\midrule

Increase the english proficiency level of this paragraph to CEFR C2 level & The function `has\_close\_elements' is designed to evaluate a list of floating-point numbers, termed `numbers', along with a specified floating-point `threshold'. Its purpose is to determine the existence of any two numbers within the list whose proximity is less than the provided threshold. The function will return 'True' if such a pair is identified; otherwise, it will return `False'. For instance, invoking `has\_close\_elements' with the list `[1.0, 2.0, 3.0]' and a threshold of `0.5' results in `False', as no two numbers in the list differ by less than `0.5' units. In contrast, calling it with `[1.0, 2.8, 3.0, 4.0, 5.0, 2.0]' and a threshold of `0.3' yields `True', given that at least one pair of numbers is separated by less than `0.3' units. \\
\midrule
 Lower the english proficiency level of this paragraph to CEFR A1 level & The function `has\_close\_elements' looks at a list of numbers called `numbers' and a number called `threshold'. It checks if any two numbers in the list are closer than the `threshold'. If yes, it says `True'. If no, it says `False'. For example, if we use the list `[1.0, 2.0, 3.0]' and `threshold' `0.5', it says `False' because no two numbers are close. But with the list `[1.0, 2.8, 3.0, 4.0, 5.0, 2.0]' and `threshold' `0.3', it says `True' because some numbers are close. \\
\midrule
Show the solution of this problem in source code. Return your response as a JSON object with one fields:
              1. ``code'': the actual Python code.  & def has\_close\_elements(numbers, threshold):    for i in range(len(numbers)):for j in range(i + 1, len(numbers)): if abs(numbers[i] - numbers[j]) \textless threshold:return True return False \# Example usage:print(has\_close\_elements([1.0, 2.0, 3.0], 0.5))  \# False print(has\_close\_elements([1.0, 2.8, 3.0, 4.0, 5.0, 2.0], 0.3)) \# True \\
\bottomrule
\end{tabular}
\caption{Increase proficiency prompt, Decrease proficiency prompt, and Generate Code prompt}
\label{tab:promptdesign2}
\end{table*}

\section{CEFR Proficiency of SE Tasks}

To answer RQ1, we first prompted the LLMs to transform each original prompt (a docstring from the HumanEval dataset) into a natural language paragraph describing the task, as shown in Table~\ref{tab:promptdesign}. This step is to understand the LLM's baseline behavior before adjusting the natural language proficiency of the description for our next step.

\textbf{CEFR for Natural Language (English).}

To determine the natural language proficiency of these generated descriptions, we utilized a CEFR word list from Kaggle~\cite{kaggle_cefrwordlist}, which contains over 10,000 English words mapped to their corresponding CEFR levels (A1 to C2). Our classification method is based on the hypothesis that a text's proficiency is defined by its most advanced vocabulary. Therefore, we assigned an overall proficiency level to each description based on the highest CEFR level of any word it contained. For example, a paragraph containing the B2-level word ``abandon'' would be classified as B2, even if all other words were A1 or A2.

\begin{table}[]
\resizebox{0.5\textwidth}{!}{%
\begin{tabular}{cccccccccc}
\toprule
\multirow{3}{*}{\textbf{CEFR Levels}} & \multicolumn{9}{c}{\textbf{HumanEval Dataset}} \\
& \multicolumn{3}{c}{\textit{GPT-4o}} & \multicolumn{3}{c}{\textit{Gemini 2.5 Pro}} & \multicolumn{3}{c}{\textit{Claude Sonnet 4}} \\
& ori. & inc. & dec. & ori. & inc. & dec. & ori. & inc. & dec. \\
\midrule
C2 & \cellcolor{gray!20}30 &79 & 1 & \cellcolor{gray!20}9 &115 & 0 & \cellcolor{gray!20}42 &143 & 1 \\
C1 & \cellcolor{gray!20}44 &67 & 12 & \cellcolor{gray!20}45 &38 & 2 & \cellcolor{gray!20}30 &19 & 10 \\
B2 & \cellcolor{gray!20}90 &18 & 151 & \cellcolor{gray!20}110 &11 & 138 & \cellcolor{gray!20}91 &2 & 144 \\
B1 & \cellcolor{gray!20}0 &0 & 0 & \cellcolor{gray!20}0 &0 & 15 & \cellcolor{gray!20}1 &0 & 8 \\
A2 & \cellcolor{gray!20}0 &0 & 0 & \cellcolor{gray!20}0 &0 & 8 & \cellcolor{gray!20}0 &0 & 1 \\
A1 & \cellcolor{gray!20}0 &0 & 0 & \cellcolor{gray!20}0 &0 & 1 & \cellcolor{gray!20}0 &0 & 0 \\
\bottomrule
\end{tabular}%
}
\caption{Distribution of CEFR Levels of 164 SE tasks as Original, Increased and Decreased proficiency problem descriptions (RQ1 and RQ2)}
\label{tab:cefr-stat}
\end{table}

Table~\ref{tab:cefr-stat} shows the distribution of the 164 problem descriptions generated by each of the three LLMs. The ``ori.'' columns represent the baseline proficiency levels relevant to RQ1. The results for the original prompts show that the LLMs predominantly generate descriptions at an intermediate proficiency level (CEFR B2) for software engineering tasks. For example, GPT-4o and Claude Sonnet 4 produced 90 and 91 descriptions at the B2 level, respectively. Notably, across all models, descriptions rated below B2 were extremely rare. Only one description (from Claude Sonnet 4) was classified as B1, and none fell into the A1 or A2 beginner levels. This suggests that a user needs at least an intermediate (B2) level of English proficiency to comfortably understand problem descriptions generated by these models.

\begin{tcolorbox}[colback=gray!5,colframe=C1blue,title= RQ1 Summary]
From the 164 problem descriptions, we find that the AI-generated descriptions proficiency is mostly intermediate level(CEFR B2) for SE problem descriptions. Interestingly, there is only one paragraph that contains only very basic words (CEFR A1- CEFR B1) from Claude Sonnet 4.
\end{tcolorbox}

\section{Prompting Proficiency of the AI-Generated Code}
\begin{figure*}
    \centering
    \includegraphics[width=0.95\linewidth]{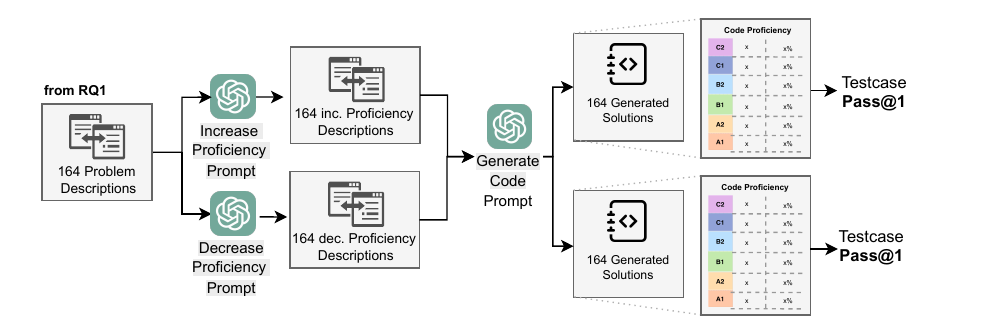}
    \caption{Evaluating Proficiency level in problem description and Proportion of code proficiency of generated
solution when using LLM (Statistic result as shown in Table~\ref{tab:cefr-stat} and Table~\ref{tab:code-stat})}
    \label{fig:rq2}
\end{figure*}

To answer RQ2, for each problem, we prompted the LLMs to rewrite the original description generated in RQ1 into two new versions: one simplified to a basic proficiency level (A1) and another elevated to an advanced level (C2). We then used these  \textit{decreased-proficiency} and  \textit{increased-proficiency} descriptions as new prompts to generate code solutions. Table~\ref{tab:promptdesign2} shows examples of these prompts. The overview of this process is depicted in Figure~\ref{fig:rq2}.

Our analysis involved two main steps: i.) Validating Prompt Proficiency: We first verified that the LLMs successfully altered the CEFR levels of the rewritten descriptions to create distinct low-proficiency and high-proficiency prompt sets. ii.) Evaluating Code Proficiency and Correctness: We then measured the proficiency of the resulting code and assessed its correctness using the pass@1 metric against the HumanEval and HumanEvalPlus test suites.

\textbf{PYCEFR for Python Code Assessment.}
To assess code proficiency, we used the reference levels from PYCEFR~\cite{pycefr2022}, a tool that classifies Python code elements into six CEFR-inspired levels (A1-C2). For example, a basic \texttt{print()} function corresponds to A1, an  \texttt{if-else} structure to B1, and an advanced construct like \texttt{zip} to C2. We analyzed each generated solution and assigned it an overall proficiency score based on the highest level code element it contained.


\begin{table}[]
\centering
\resizebox{0.5\textwidth}{!}{%
\begin{tabular}{ccccccc}
\toprule
\multirow{3}{*}{\textbf{Code Proficiency}} & \multicolumn{6}{c}{\textbf{HumanEval Dataset}} \\
& \multicolumn{2}{c}{\cellcolor{gray!20}\textit{GPT-4o}} & \multicolumn{2}{c}{\textit{Gemini 2.5 Pro}} & \multicolumn{2}{c}{\textit{Claude Sonnet 4}} \\
& \cellcolor{gray!20}inc. & \cellcolor{gray!20}dec. & inc. & dec. & inc. & dec. \\
\midrule
C2 & \cellcolor{gray!20}17 & \cellcolor{gray!20}17 & 15 & 9 & 7 & 7 \\
C1 & \cellcolor{gray!20}24 & \cellcolor{gray!20}25 & 13 & 14 & 14 & 14 \\
B2 & \cellcolor{gray!20}20 & \cellcolor{gray!20}15 & 11 & 1 & 10 & 9 \\
B1 & \cellcolor{gray!20}69 & \cellcolor{gray!20}41 & 24 & 27 & 20 & 21 \\
A2 & \cellcolor{gray!20}18 & \cellcolor{gray!20}26 & 47 & 53 & 63 & 60 \\
A1 & \cellcolor{gray!20}16 & \cellcolor{gray!20}40 & 54 & 60 & 50 & 53 \\
\midrule
\multicolumn{7}{c}{\textbf{Statistical Test Result}} \\
\midrule
p-value &  \multicolumn{2}{c}{\cellcolor{gray!20}0.001**} & \multicolumn{2}{c}{0.057*} & \multicolumn{2}{c}{0.998} \\
CramerV &  \multicolumn{2}{c}{{0.244}} & \multicolumn{2}{c}{0.180} & \multicolumn{2}{c}{0.027} \\
Effect size &  \multicolumn{2}{c}{{small}} & \multicolumn{2}{c}{small} & \multicolumn{2}{c}{small} \\
\midrule
\multicolumn{7}{c}{\textbf{Pass@1 Benchmark Result}} \\
\midrule
HumanEval & 79.3\% & 76.2\% & 87.2\% & 71.3\% & 93.9\% & 91.5\% \\
HumanEval+ & 76.2\% & 70.7\% & 83.5\% & 69.5\% & 88.4\% & 85.4\% \\
\bottomrule
\end{tabular}
}
\caption{Distribution of Code Proficiency in Increased and Decreased proficiency prompt for 164 SE-tasks in HumanEval and Pass@1 result (RQ2).}
\label{tab:code-stat}
\end{table}

\subsection{Analysis 1 - Verifying the CEFR proficiency of descriptions}
As shown in Table~\ref{tab:cefr-stat}, our prompting strategy successfully altered the natural language proficiency of the problem descriptions. For the increasing proficiency task, all models produced a large number of C1 and C2-level descriptions, confirming they could effectively elevate the language. Conversely, for the decreasing proficiency task, models were less effective at simplification. While some descriptions were lowered to B1, A2, or A1, a significant portion remained at the B2 (Intermediate) level, indicating a floor effect in the models' ability to simplify technical language.

\subsection{Analysis 2 - Evaluating the impact on the code proficiency of generated solutions}
\indent \textit{Code Proficiency:}
Table~\ref{tab:code-stat} presents the distribution of code proficiency for solutions generated from the increased and decreased-proficiency prompts. We conducted a Chi-Square test of independence to determine if the natural language proficiency level had a statistically significant effect on the code proficiency level. \\
\indent Our findings vary by model. For GPT-4o, the test yielded a statistically significant result (p$<$0.05), rejecting the null hypothesis and indicating that the natural language proficiency was associated with differences in the distribution code proficiency. For Gemini 2.5 Pro, the result approached significance (p=0.057), suggesting a potential but less pronounced relationship. In contrast, Claude Sonnet 4 showed no significant effect (p=0.998), implying the generated code from its was independent of the CEFR natural language proficiency. These results suggest that for some LLMs, adjusting natural language proficiency could alter the code proficiency of the generated solution, both increasing and decreasing proficiency. \\
\indent \textit{Code Correctness:}
The impact on code correctness was more uniform across all models. As shown in the Pass@1 Benchmark Result section of Table~\ref{tab:code-stat}, solutions generated from increased-proficiency prompts consistently achieved higher pass@1 scores on both HumanEval and HumanEvalPlus benchmarks. For every model, the decreased-proficiency prompts led to a noticeable drop in correctness, suggesting that richer, more advanced language enables the models to generate more reliable solutions.
\vspace{-0.22cm}
\begin{tcolorbox}[colback=gray!5,colframe=C1blue,title= RQ2 Summary]
Adjusting (increasing or decreasing) the natural language proficiency of the SE task description resulted in the code with different proficiencies for GPT-4o but not Gemini 2.5 Pro and Claude Sonnet 4. However, the generated solution from increased natural language proficiency description tends to achieve higher pass@1 test coverage than the generated solution from decreased natural language proficiency description of three LLMs.
\end{tcolorbox}

\section{Limitations}

This section outlines the key limitations of our study and potential threats to the validity of our findings.

\textbf{Prompt Design.}
One limitation stems from the prompt design used to manipulate natural language proficiency. While we aimed to standardize prompts and control for content, subtle variations in phrasing, tone, or complexity may have influenced the quality of the generated responses.

\textbf{Proficiency Assessment.} Our approach to assessing both natural language and code proficiency relies on a consistent hypothesis: the presence of advanced elements serves as a strong proxy for overall proficiency. For natural language proficiency, our analysis focuses on the vocabulary within a given text, rather than its grammatical structure or other syntactic aspects. Similarly, for code proficiency, we evaluate the sophistication of the solution based on the specific code constructs and elements it contains. This method provides a focused, rather than holistic, measure of proficiency. However, this approach is grounded in the principle that encountering a single, critical, and unfamiliar element can significantly impede comprehension. For instance, a single unknown yet crucial word can obscure the meaning of an entire paragraph. Likewise, in programming, an unfamiliar function or syntactic structure can prevent a developer from fully understanding the code's intent and functionality. Therefore, the implications of our findings should therefore be understood primarily in the context of construct complexity and its impact on immediate user comprehension. A more comprehensive proficiency model could integrate our construct-based analysis with established software engineering metrics, such as cyclomatic complexity, readability scores, and static analysis for code smells, to provide a richer understanding of AI-generated software quality.

\textbf{Task Scope.}
Our study is based on the HumanEval benchmark dataset, which focuses on short, self-contained Python programming tasks with a heavy emphasis on algorithmic reasoning. As such, the results may not generalize to broader or more complex domains of software engineering, such as API integration, system design, large-scale code generation, or debugging tasks that involve maintaining context over longer sequences.

\textbf{Model Dependency.}
All results in this study were obtained using three LLMs (GPT-4o, Gemini 2.5 Pro, and Claude Sonnet 4). While this allows for controlled experimentation, it also limits the generalization of our findings. Other LLMs especially those trained on different data or optimized with different objectives may respond differently to variations in natural language proficiency.


\section{Implications and Future Outlook}

\textbf{For Developers}
Findings from RQ2 demonstrate that the natural language proficiency of prompts could significantly influence the quality and sophistication of code generated by large language models. Developers using AI-assisted code generation tools should therefore be mindful not only of the technical accuracy and completeness of their prompts but also of their natural language clarity and proficiency. Well-structured, high-proficiency language in prompts can lead to the generation of more advanced, correct, and functional code. This insight encourages developers to view prompt engineering as a multi-dimensional practice involving both contextual clarity and language precision. However, this sensitivity is not universal across all models. A development team, particularly one with diverse levels of natural language proficiency (e.g., a mix of native and non-native speakers), might achieve more consistent and predictable results by choosing an LLM that is less sensitive to variations in natural language expression. This ensures that all team members can generate reliable code regardless of their natural language style.

\textbf{For Non programmers}
The result from RQ1 reveals that large language models can effectively translate structured code prompts into natural language descriptions. This capability opens the door for non-programmers to better understand software tasks and participate in software-related conversations. However, our results also indicate that these descriptions, even after simplification, tend to remain at a minimum B2 CEFR level. This suggests a potential barrier to accessibility: individuals with lower English proficiency may still struggle to fully grasp software concepts conveyed by large language models. 

\textbf{For Researchers}
This study contributes to the emerging body of research examining the intersection between natural language proficiency and code proficiency. Our findings underscore the significance of natural language factors in prompt-based programming and highlight the measurable influence of language quality on the proficiency of code output. The relationship between natural language and code proficiency, as explored through RQ2, provides a foundation for future research on educational tools, accessibility in programming, and improvements in prompt engineering. Moreover, these results suggest promising directions for designing more inclusive AI systems that adapt to users natural language capabilities while maintaining technical accuracy.

\textbf{Future Outlook}
It is now clear that AI-assistants and FM-powered tools are here to stay. Therefore,  to embrace the technology, we also need to understand how to control not only through the precision of the prompt but also the proficiency and nuances behind the natural language used. We envision that these results lay the groundwork for future research directions into more predictable refining of AI-generated code, assessing developer skills and ultimately the quality of solutions to software engineering tasks. 

\section{Acknowledgement}
This work is supported by the JSPS KAKENHI Grant Number JP24H00692 and JP23K28065.

\bibliographystyle{IEEEtran}
\bibliography{ref}
\vspace{1\baselineskip}

{\setlength\intextsep{0pt}
\begin{wrapfigure}{l}{25mm} 
    \includegraphics[width=1in,height=1.25in,clip,keepaspectratio]{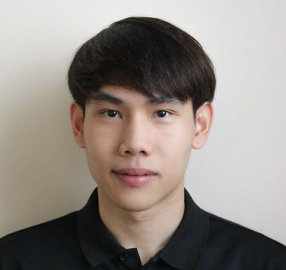}
\end{wrapfigure}\par
\noindent\textbf{Ruksit Rojpaisarnkit} is a PhD student in the Software Engineering Laboratory at the Graduate School of Science and Technology, Nara Institute of Science and Technology (NAIST). His main research interests were related to the human aspect of software engineering, code proficiency, and data mining.
Find him at \url{https://www.linkedin.com/in/ruksit-rojpaisarnkit-76b72417a/}.
\par}

{\setlength\intextsep{0pt}
\begin{wrapfigure}{l}{25mm} 
    \includegraphics[width=1in,height=1.05in,clip]{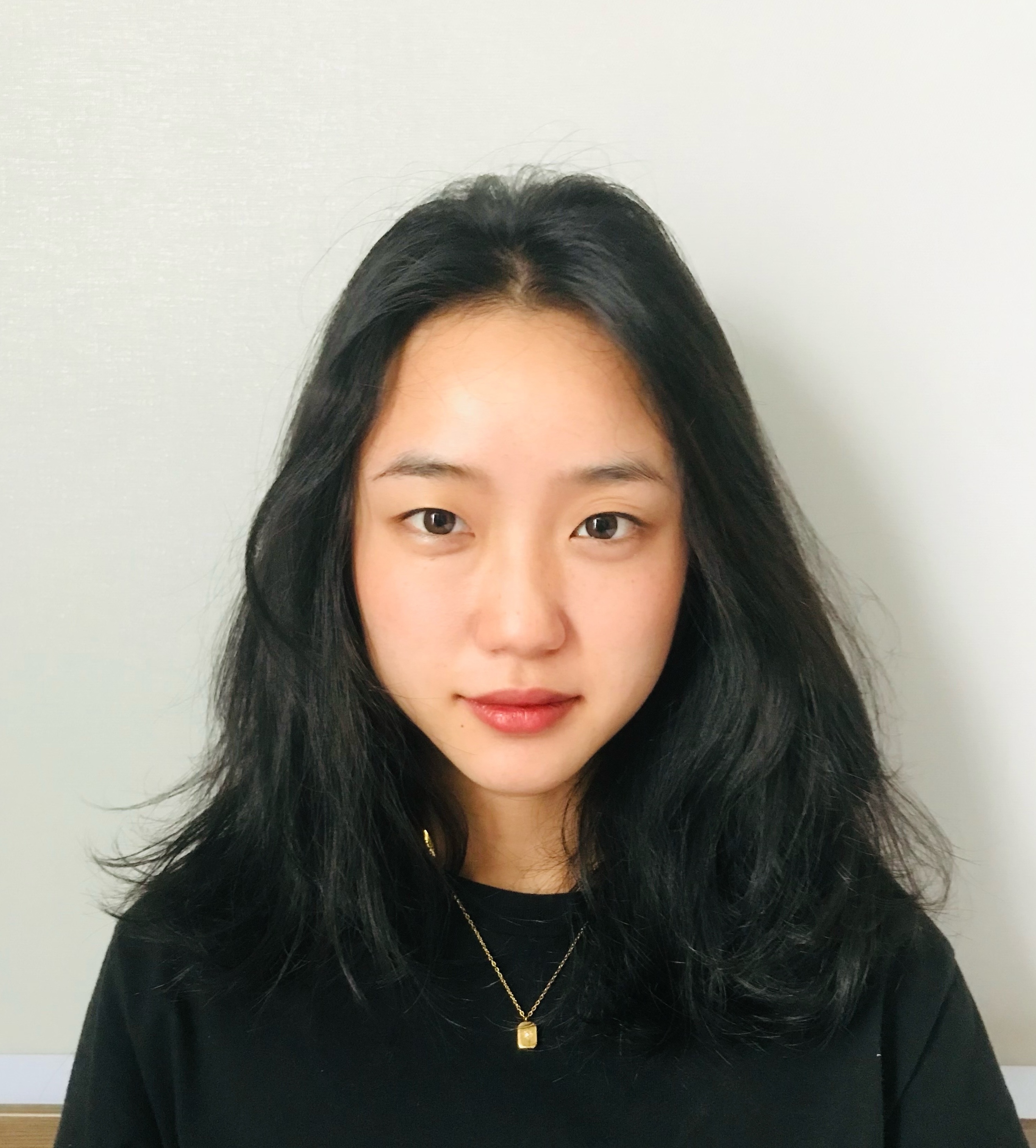}
\end{wrapfigure}\par
\noindent\textbf{Youmei Fan} is an Assistant Professor in the  Software Engineering Laboratory at the Graduate School of Science and Technology, Nara Institute of Science and Technology (NAIST). She completed a doctoral course in Software Engineering Laboratory at the Graduate School of Science and Technology, Nara Institute of Science and Technology (NAIST), Japan. During her doctoral degree, her main research interests were related to the human aspect of software engineering, open-source software sustainability, and data mining.
Find her at \url{https://www.linkedin.com/in/youmei-fan-513a161a6/}.
\par}

{\setlength\intextsep{0pt}
\begin{wrapfigure}{l}{25mm} 
    \includegraphics[width=1in,height=1.05in,clip]{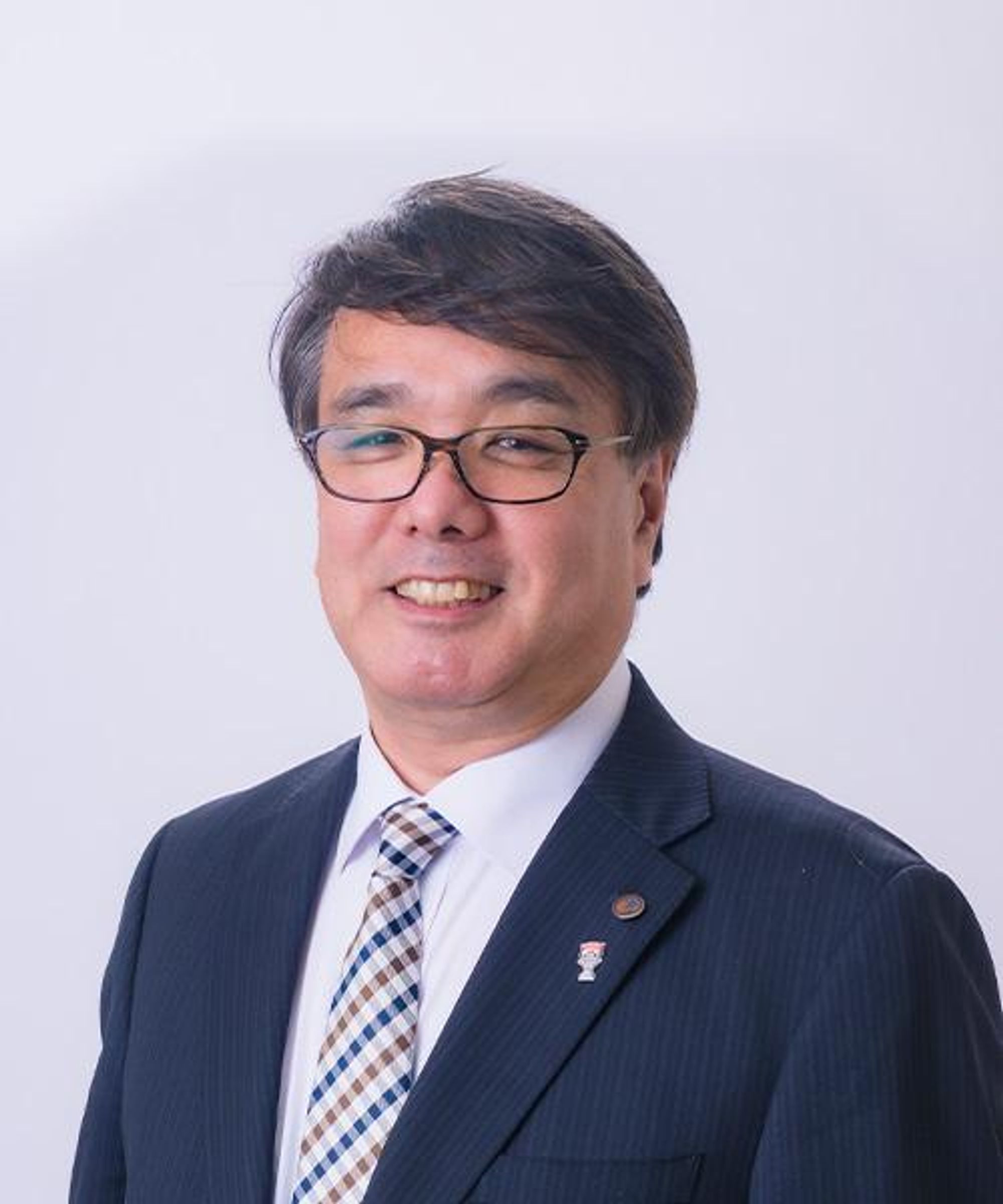}
\end{wrapfigure}\par
\noindent\textbf{Kenichi Matsumoto} is a Professor in the Software Engineering Laboratory at the Graduate School of Science and Technology, Nara Institute of Science and Technology (NAIST) Contact him at matumoto@is.naist.jp. \par}

{\setlength\intextsep{0pt}
\begin{wrapfigure}{l}{25mm} 
    \includegraphics[width=1in,height=1.05in,clip]{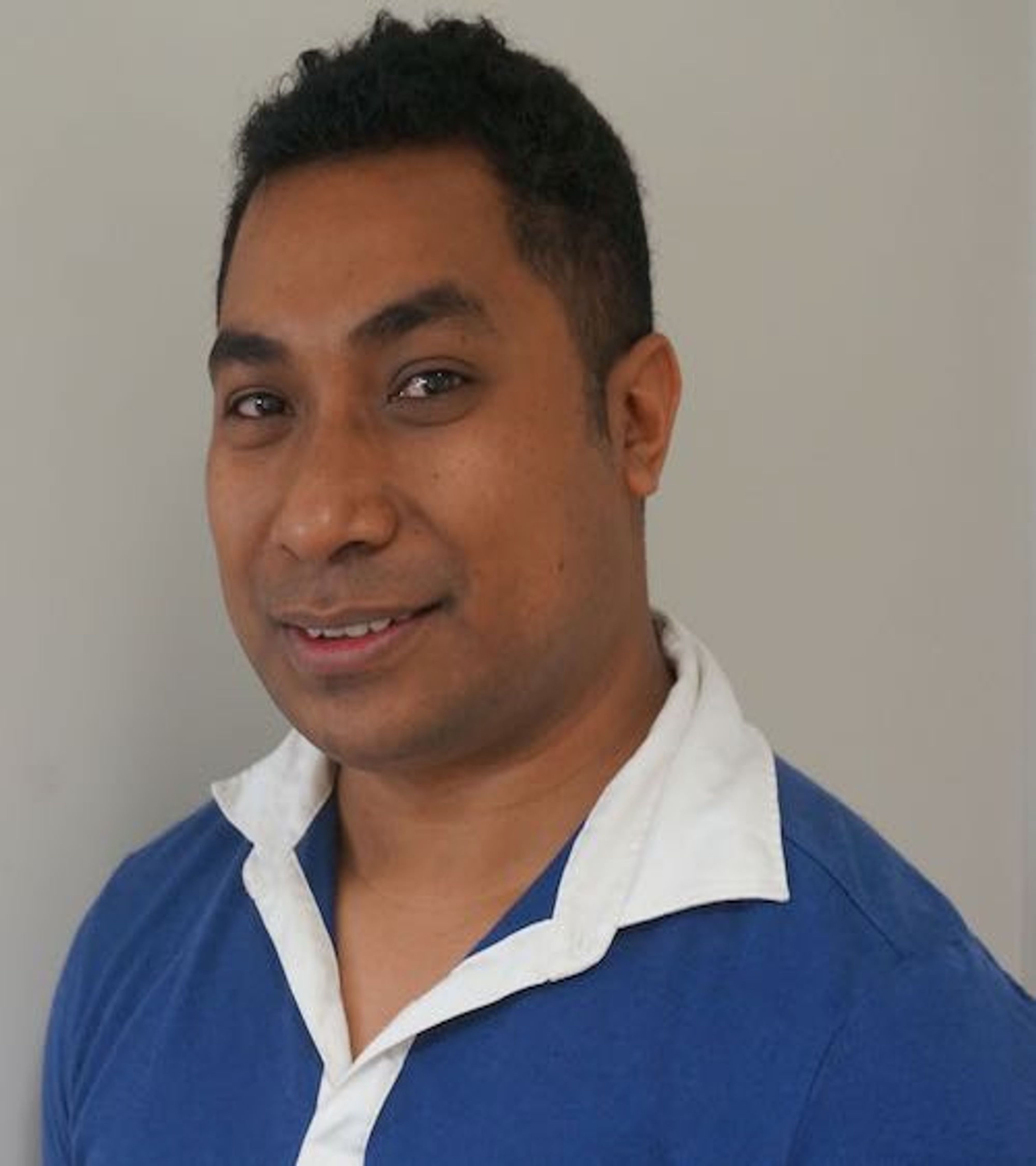}
\end{wrapfigure}\par
\noindent\textbf{Raula Gaikovina Kula} is a Professor at The University of Osaka. He received his Ph.D. from Nara Institute of Science and Technology (NAIST) in 2013, later joining as a Specially-Apppointed Assistant Professor from (2013-2016) at Osaka University. He then continued as a Specially-Appointed Assistant Professor from (2017), later becoming an Assistant Professor (2017-2023), and a Associate Professor (2023-2024) at NAIST.  \par}

\end{document}